\documentclass[aps,prl,twocolumn,preprintnumbers,groupedaddress,superscriptaddress,floatfix,tightenlines,reprint]{revtex4-1}
\usepackage{mathrsfs,natbib}
\usepackage{graphics,epsfig,color, graphicx}
\usepackage{verbatim}
\usepackage{bm}

\usepackage{graphicx,epsf,amssymb,bbm,amsbsy,amsfonts,amssymb,amsmath}
\usepackage{hyperref}

\hypersetup{
    colorlinks=true,       
    linkcolor=red,          
    citecolor=blue,        
    filecolor=magenta,      
    urlcolor=blue           
}

\usepackage{amsmath,amssymb,multirow,graphicx}
\usepackage{centernot}
\usepackage{color}
\allowdisplaybreaks[1]
\usepackage{tikz}
 \tikzset{node distance=2cm, auto}
\usepackage{bbm} 
\usepackage{youngtab}

\renewcommand{\Im}{\text{Im }}

\def\Im{\text{Im}}

\def\tr{\text{tr}}

\def\a{{\alpha}}

\def\g{{\gamma}}
\def\G{{\Gamma}}

\def\f{{\varphi}}

\def\*{\star}
\def\Z{{\mathbb Z}}
\def\R{{\mathbb R}}
\def\coeff#1#2{{\textstyle {\frac {#1}{#2}}}}

\def\half{\coeff 12}

\usepackage{verbatim}   
\usepackage[all]{xy}
\usepackage{bm}  
\def\Dslash{{\rlap{\raise 1pt \hbox{$\>/$}}D}}
\def\Pslash{{\rlap{\raise  1pt \hbox{$\>/$}}\,\partial}}

\newcommand{\V}{{\mathcal{V}}}

\newcommand{\be}{\begin{equation}}      
\newcommand{\ee}{\end{equation}}      
\newcommand{\bea}{\begin{eqnarray}}      
\newcommand{\eea}{\end{eqnarray}}

\begin{document}


\title{Cluster expansion and resurgence in Polyakov model}
\author{Cihan Pazarba\c{s}{\i}}
\affiliation{Physics Department, Bo\u{g}azi\c{c}i University
	34342 Bebek, Istanbul, Turkey}
\affiliation{Department of Physics, North Carolina State University, Raleigh, NC 27695, USA}
\author{Mithat \"Unsal}
\affiliation{Department of Physics, North Carolina State University, Raleigh, NC 27695, USA}
\email{unsal.mithat@gmail.com}

\begin{abstract}
In Polyakov model, a non-perturbative  mass gap is formed at leading order semi-classics by   instanton effects.  By using the notions of critical points at infinity, cluster expansion and Lefschetz thimbles, 
we show that a third order effect  in semi-classics gives  an imaginary ambiguous contribution to mass gap, which is  supposed to be real and unambiguous.   This  is  troublesome for the original  analysis, and it 
%
  is difficult   to resolve this issue directly in QFT.    However, 
we find a new compactification of Polyakov model to quantum mechanics, by using a background 't Hooft flux (or coupling to TQFT).   The compactification has the merit of remembering  the  monopole-instantons of the full QFT  
within  Born-Oppenheimer (BO) approximation, while the periodic compactification does not.  In  QM,  we prove the resurgent cancellation of the ambiguity in 3-instanton sector against ambiguity in the Borel resummation of the perturbation theory around  1-instanton. 
 Assuming that this  result holds in QFT,  we provide a  large-order asymptotics of  perturbation theory around  
 perturbative vacuum and instanton.   
 
\end{abstract}


\maketitle



\noindent
{\bf Introduction:} Polyakov model is a prototypical example of   non-perturbatively calculable weakly coupled quantum field theory \cite{Polyakov:1976fu}. It is  
by now standard textbook material in QFT, condensed matter physics and it is also intimately tied with  statistical field theory 
 of Coulomb gases \cite{Deligne:1999qp, Banks:2008tpa, fradkin_2013, Shifman201201}.
Despite the fact that some fundamental facts about the theory are known for more than four decades now,  after the advent of resurgence  \cite{Dunne:2012ae, Cherman:2014ofa, Dunne:2016nmc} and Lefschetz thimbles \cite{Witten:2010cx}
 many subtle issues  emerge concerning  this  and other calculable theories.  An important issue is following. It is known that the mass gap in the theory is sourced by monopole-instantons on $\R^3$, and is of order $m_g^2 \sim e^{-S_0}$ where $S_0$ is monopole-instanton (${\cal M}_{\alpha_i}$) action,  with magnetic charge $\alpha_i \in \Delta^{0}$ in the simple root system. How one does incorporate saddles with higher action 
$(nS_0, n \geq 2)$? Should one care about them? Do they contribute to mass gap?  
This class of questions are usually not addressed and swept under the rug by assuming  that these are higher order quantitative corrections, and not important  \cite{Polyakov:1976fu,  fradkin_2013}, even 
in more mathematical treatments of the theory \cite{Deligne:1999qp}. 
  Of course, there are  also theories  in which mass gap is induced at higher order  effect in monopole-expansion  due to Berry phase  \cite{Read:1990zza} or  topological theta angle \cite{Unsal:2012zj} induced destructive interference between leading monopole events,  or as an effect of index theorem \cite{Unsal:2007jx}. But here 
we address  the above  questions in simple Polyakov model, where higher order effect just seem like nuissance, by using the concepts of critical point at infinity, quasi-zero mode Lefschetz thimbles  and cluster expansion \cite{kardar} systematically.

Here is the main point of our analysis. The mass gap in semi-classical expansion in Polyakov model is  of the following form: 
(ignoring inessential factors to lessen the clutter)
\begin{align}
(m_g^2)_{\pm} &\sim \left( e^{- S_0}  P_{1}   +   e^{- 2 S_0}  P_{2} +   e^{- 3 S_0} P_{3} +
 \ldots  \right)     \cr
&\pm i \left(   e^{- 3S_0} + \ldots \right) 
\label{P-Em}
\end{align}
where $P_i$ denotes perturbative expansions around relevant saddle. 
It is reasonable to drop  $O(e^{- 2 S_0} )$  terms in the real part of his analysis, as they provide only minor quantitative corrections. But as we emphasize, there is a  new effect in third order in semi-classics, which renders the semi-classical expansion multi-fold ambiguous and void of meaning.  It is actually not correct to ignore $\pm i \left(   e^{- 3S_0} \right)$ because it is an  effect of  different nature, giving mass an imaginary ambiguous part.   Therefore, one is entitled to ask whether Polyakov's analysis is rigorous enough even within semi-classics? In particular, for the famous result on mass gap to be justified,  one needs   
a mechanism for the cancellation of imaginary ambiguity on $\R^3$. 


  
The type of ambiguities that appear in \eqref{P-Em}  is in fact expected.   The reason for this is because ${\rm Arg} (\hbar) = 0$ and   ${\rm Arg} (\hbar) = \pi$  are  in general Stokes lines.  On Stokes lines,  contributions of a sub-class of saddles can indeed be multi-fold ambiguous.  In fact,  there are infinitely many critical points at infinity, and  generically, there are multi-fold ambiguities.  
 It is desirable to resolve these pathological features in order to  make Polyakov's solution meaningful in this new light.

According to resurgence,  there is another ambiguity in the story. Perturbation theory around perturbative vacuum saddle, one-instanton saddle etc are all expected to be divergent asymptotic expansions \cite{Bogomolny:1980ur,Zinn-Justin:1981qzi, 
 Dunne:2014bca, Alvarez,Fujimori:2017oab}.   They are also expected to be non-Borel summable, meaning that Borel resummation  of perturbation theory  around each saddle is  multifold  ambiguous. Resurgence implies that for the theory to be meaningful, these two types of ambiguities must cancel  around each  sector of the theory.   However, demonstrating this in a generic QFT  is hard. It is possible to take mileage on this problem using the idea of adiabatic continuity and turning on background 
fields   \cite{Dunne:2012ae, Cherman:2014ofa, Dunne:2016nmc}, 
  by working with QFTs with special properties such as integrability \cite{Marino:2019eym, Marino:2021six} or working with rather special QFTs in which one has  a good knowledge of perturbation theory \cite{Gukov:2016njj}. 
     In this work, we employ a  't Hooft flux  background (couple the theory to a TQFT background)  to tackle this problem in Polyakov model. 

 \vspace{0.3cm}
 \noindent
{\bf Basic:} Polyakov model  is given as an    $SU(N)$ non-abelian gauge theory coupled to an adjoint scalar field in a 3d Euclidean 
space:  
    \begin{equation}\label{PolyakovModel_Action}
    	S = \int d^3x   \; \frac{1}{2 g_3^2}\bigg(\tr F^2  + \tr (D \f )^2   +  \lambda  V(\f)  \bigg) 
    \end{equation}
where the potential  $V(\f)$  leads to the abelianization of gauge dynamics down to $U(1)^{N-1}$. We assume without loss of generality that the eigenvalues of $\phi$ are uniformly separated, $v_i - v_{i+1} = v$.  
In the   BPS limit, ($\lambda \ll 1$), the theory has saddles which are solutions to  (anti-)self-duality equations
$	F  = \pm \*_3 D\f $ with topological (magnetic)  charges 
$	Q_{M_i} = \frac{2\pi}{g_3}\a_i $ and actions 
$  	S^{(i)}_0  	=   \frac{4 \pi v}{g_3^2}  \equiv   \frac{{\rm s}_0} {g^2}$, where  $\alpha_i \in \Delta^{0}$ are $N-1$ simple roots, where $g^2 = g_3^2/v$ is a dimensionless expansion parameter.    
 The monopole operators are 
$ {\cal M}_{\alpha_i} \sim   (S_0)^2 e^{-S_0} e^{ - \frac{4 \pi}{g^2} \tilde {\bm \phi} (x) \cdot {\bm \alpha}_i + i \bm{\alpha}_i \cdot \bm{\sigma}(x)}$, where $  \tilde {\bm \phi} (x), \sigma(x)  $  are fluctuation of Cartan components of adjoint scalar and dual photon,  
$(S_0)^2$ arise from the four zero modes of the monopole. 
The former can be set to zero in the description of the long distance theory. 
The proliferation of monopoles generates a mass gap for gauge fluctuations as discovered by Polyakov \cite{Polyakov:1976fu}, see  also
\cite{Deligne:1999qp,  Anber:2013xfa}.
 
\begin{figure}[t]
\begin{center}
\includegraphics[width = 0.155 \textwidth]{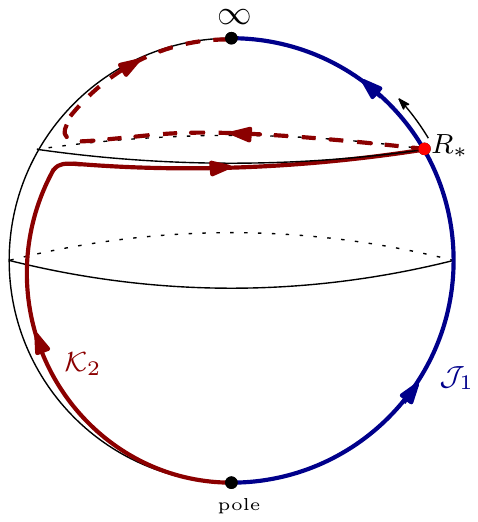}
\includegraphics[width = 0.155 \textwidth]{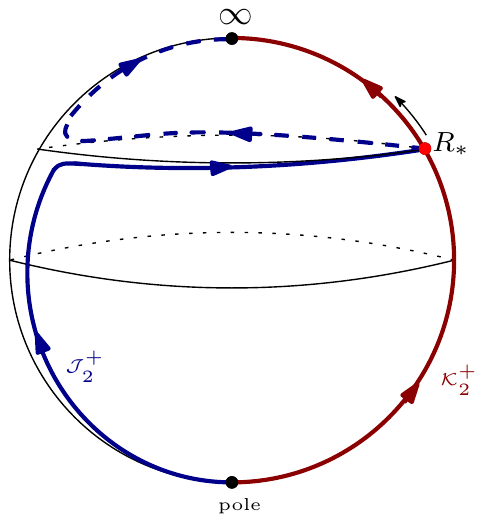}
\includegraphics[width = 0.155 \textwidth]{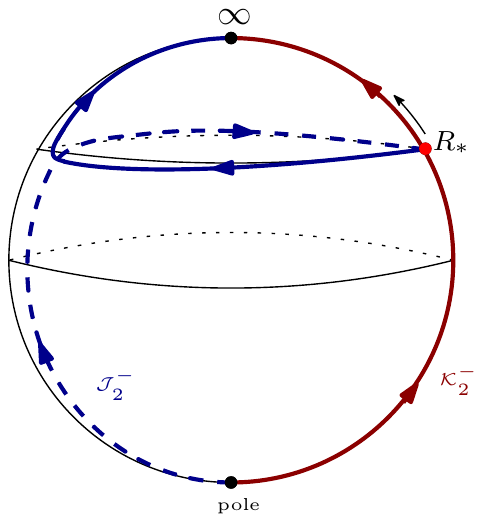}
\caption{ ${\cal J}_{1}$ and  ${\cal J}_{2}^{\pm}$   are  the (regularized) steepest descent cycles for repulsive  and 
attractive interaction.   The former is unique, and  
the latter  is two-fold ambiguous at   ${\rm arg}(g^2) = 0$.  
The point $R^*$ tends to infinity as the cut-off is removed,  but the loops at infinity remains. 
 }
 \vspace{-0.7cm}
\label{fig:cycles}
\end{center}
\end{figure}

 \vspace{0.3cm}
 \noindent
{\bf Critical points at infinity and cluster expansion:} The model, apart from regular saddles, also possesses critical points at infinity \cite{ Nekrasov:2018pqq, Behtash:2018voa, Witten}.  These critical points  are very likely one of the most important concepts  in semi-classics, yet, there is a minuscule amount of work on systematizing them, or a heavy  burden of misunderstandings emanating from 70s related to them, see 
\cite{ Nekrasov:2018pqq, Behtash:2018voa}. 
These configurations, up to our knowledge, are not addressed at all in the context of Polyakov model,  and will play important role below. 

Consider a monopole-monopole or monopole-antimonopole pair. The interaction between the two in the BPS $(\lambda \rightarrow 0)$ limit is:
\begin{align}
 	V_{\mathrm{int}}(r) =  \left\{  \begin{array}{ll} 
	   0 & \qquad {\rm for}  \;   ({\cal M}_{i},   {\cal M}_j)   \cr
    \frac{2 \pi}{g^2} \frac{\a_i \cdot (- \a_j) }{r},    & \qquad {\rm for}  \;   ({\cal M}_i,   \overline  {\cal M}_j) 
\end{array} \right.
\end{align}
 Thus,  $(\cal M,   {\cal M})$  do not interact \cite{Manton:1977er, Affleck:1982as},  but  $(\cal M,   \overline  {\cal M}) $ pairs interact, attractively for $i=j$ and repulsively for $j = i \pm 1$.  At any finite separation, since  
$ V_{\rm int}' |_{r < \infty} \neq 0 $,\
 these pairs are not  exact solutions. But at $r= \infty$, they become exact solutions, hence  the name.  Such configurations are   genuine critical points, but they  are non-Gaussian, i.e.  $V_{\rm int}''|_{r= \infty} =0$, unlike Gaussian saddles.   Because of this property,  one needs to integrate over the whole steepest descent cycle to find the effect of such pairs. 
 
 The integrals that give the contribution of the second order effects in  semi-classics  are of the form: 
  \begin{align}\label{PartitionFunction_TwoInstanton}
    	Z_2  &= [{\cal M}_i  ]  [ \overline {\cal M}_{j}  ]  \int {\rm d^3r_1  d^3r_2}  \, e^{- V_{\mathrm{int}}(| \bf r_1 -  r_2|)} 
	    \end{align} 
The measure can be expressed in terms of center-of-action coordinate, the integral over which gives a space-time volume factor $\V$, and the relative coordinate, which corresponds to quasi-zero mode direction: 
$[{\cal M}_i  ]  [ \overline {\cal M}_{j}  ]   {\cal V}\, {4 \pi} \int  dr r^2    e^{- V_{\mathrm{int}}(r)} $. 
 This type of integrals appears in the standard cluster expansion in statistical field theory \cite{kardar}.

 In  semi-classics, steepest descent cycles are not necessarily real.  Let us call $r \rightarrow z \in \mathbb C$. 
   The steepest descent cycles  can end up  at points where  $e^{- f(z)/g^2}  \rightarrow 0$.  For polynomial $f(z)$, this gives a homology cycle decomposition of the integration \cite{Witten:2010cx}, in terms of cycles that ends at infinity in certain  wedges. 
   For the Coulomb potential, the end point is a pole at $z=0$  similar to \cite{Basar:2013eka}, and the critical point is at $z=\infty$. 
   For    
$e^{- 1/(g^2 z)  }$, the cycle must enter to the pole in $\arg(z)=0$ direction, while  $e^{+ 1/ (g^2 z) }$, the cycle  must  enter to the pole  in $\arg(z)=\pi$ direction for steepest descent. So, the steepest descent directions  for the attractive and repulsive potentials   are  different.

 Steepest descent cycles are easiest to visualize if we map  $r \equiv z  \in \mathbb C$ complex domain to Riemann sphere by using one-point compactification.  To see the thimbles more clearly, we can introduce a  regulator for the  integral, find the steepest descent, and ultimately remove the regulator. For the two interaction types, the 
  descent cycles for $\arg (g^2) = 0^{\pm} $  are given by
 \begin{align} 
& {\cal J}_1 (0):   z \in [0, \infty], \qquad {\rm repulsive}, \cr
& {\cal J}_2 (0^{\pm}):  z  \in [0, -\infty] \cup [ C_{\infty}^{\circlearrowleft} \; { \rm or} \; C_{\infty}^{\circlearrowright} ] 
\qquad {\rm attractive},
\end{align}
and are shown in Fig.\ref{fig:cycles}. 
The integrations  are  given by 
     \begin{align} 
     \label{subext}
      	Z_2  & \sim   \xi^2  \V^2     \qquad   \qquad \qquad \qquad  \;\;\; {\rm non-interacting}, \cr
   	Z_2  & \sim   \xi^2  \V  \left( \V+           I(g^2) \right)     \qquad  \qquad {\rm repulsive}, \cr
   	Z_{2, \pm}  & \sim   \xi^2  \V  \left( \V+           I(g^2 e^{\pm i \pi}  )\right)   
\qquad {\rm attractive},
\end{align}
 where the extensive part corresponds to free (non-interacting)  monopole gas with fugacity $\xi$ and $I(g^2)$ in the   sub-extensive term is called second virial coefficient, capturing the effect of interactions. It is  given by: 
  \begin{align}
    	\label{C2m}
    I(g^2) =     	\frac{4 \pi}{6}   \Big(\frac{2 \pi |\a_i\cdot\a_j|}{g^2}\Big)^{3}   \Big(  \ln \Big(    \frac{2 \pi |\a_i\cdot\a_j| }{g^2 }     \Big) + \g -   \frac{11}{6}   \Big)  .
    \end{align}
  In the repulsive case, subextensive part  is usually called magnetic bion, and its amplitude is  
   $[{\cal M}_{\alpha_{i}}  \overline {\cal M}_{\alpha_{i\pm1}} ]=    I(g^2) [{\cal M}_{\alpha_{i}}] [ \overline {\cal M}_{\alpha_{i\pm1}} ]   $ \cite{Argyres:2012ka}.     In statistical physics, this is a configuration in  2-cluster, ${\cal C}_2$.  
   
   Attractive case is  more interesting. First, note that 
     \begin{align}
    	\label{C2n}
    I(g^2  e^{\pm i \pi}) =   e^{i \pi} I(g^2)  \pm i     	\frac{2 \pi^2}{3}   \Big(\frac{2 \pi |\a_i\cdot\a_j|}{g^2}\Big)^{3}  ,
        \end{align}
   which implies two different remarkable phenomena for this element of  ${\cal C}_2$.  First, we identify the sub-extensive part in the attractive case with neutral bions,  
    $[{\cal M}_{\alpha_{i}} \overline {\cal M}_{\alpha_{i}} ]_{\pm}=    I(g^2 e^{\pm i \pi} ) [{\cal M}_{\alpha_{i}}] [ \overline {\cal M}_{\alpha_{i}} ]   $ \cite{Argyres:2012ka}.  
   First, the contribution is two-fold ambiguous. This is expected, because we are formulating the path integral on a Stokes line, and the configurations with attractive interactions are  expected to lead to  two-fold ambiguous results. At least in some limit of QFT, we will prove that this    two-fold ambiguity cancels against another ambiguity. 
   
   The overall phase in front of \eqref{C2n} is equally  interesting. It tells us that the fugacity of the two-cluster elements  
   can be complex! 
   \begin{align}
   {\rm Arg} ([{\cal M}_{\alpha_{i}}  \overline {\cal M}_{\alpha_{i\pm1}} ])=    {\rm Arg} ( {\rm Re} [{\cal M}_{\alpha_{i}}  \overline {\cal M}_{\alpha_{i}} ]_{\pm}) + \pi
   \label{bions}
\end{align} 
This is in some sense similar to   \cite{Read:1990zza, Unsal:2012zj} where there is a relative topological  phase (sourced by Berry phase or $\theta$-angle) between monopole-events. 
This relative phase between the two contributions is now sourced by the phase associated with thimble and is called  hidden topological angle (HTA) \cite{Behtash:2015kna}. It is known to play crucial role in semi-classics. For example,  in ${\cal N}=1$ on $\R^3 \times S^1$,   and  SUSY and QES quantum mechanics,   the vanishing of the vacuum energy  (or NP contributions)   is  due to  relative phase between these two types of events.  For other aspects of  bions, see  \cite{Dunne:2016nmc, Anber:2011de, Poppitz:2011wy, Unsal:2007jx, Fujimori:2017osz, Misumi:2016fno, Misumi:2014jua, Morikawa:2020agf, Fujimori:2017oab, Misumi:2014bsa, Fujimori:2018kqp}.

%

At third  order in semi-classics,  events such as  $[{\cal M}_{\alpha_{i}}  \overline {\cal M}_{\alpha_{i}}  {\cal M}_{\alpha_{i}}]_{\pm} = O(e^{-3S_0})  \pm i  O(e^{-3S_0}) $  provide a  two-fold ambiguous contribution to mass gap, which is supposed to be real and unambiguous.  The fact that the  ambiguity in mass  first appears in the third order comes from the structure of resurgence triangle \cite{Dunne:2012ae}.  As it stands, this is quite disturbing for  Polyakov's well-known solution. 

We can anticipate that these ambiguities in  2-event and 3-event  contributions must be related  with the non-Borel summability of the perturbation theory around  perturbative vacuum and 1-instanton sector, respectively.  The left/right Borel resummation  is two-fold ambiguous as well, and these two types of ambiguities  are expected to cancel. 

However, it is difficult   to test this scenario in full QFT.  As we describe, it is also not possible to address this question by using a naive reduction  of QFT to QM  within Born-Oppenheimer (BO) approximation. However, we propose a compactification with discrete  't Hooft flux in which monopole actions remain the same, and  such cancellations in QM limit can be shown. 

 \vspace{0.3cm}
 \noindent
{\bf Periodic $T^2$ compactification does not work: } One may consider compactification on $T^2 \times \R$, and study the interplay of instantons and perturbation theory in  QM  reduction. However, a problem awaits us here.  The states in the QM are described in terms of magnetic flux  
 through $T^2$,  $ | \Phi \rangle$. The lowest energy state is  zero flux state  $ |\bf 0 \rangle$  and   magnetic flux  states with  non-zero flux   $ |\bm  \alpha_a \rangle$  are parametrically separated in  energy: 
  $E_0= 0$ and  $E_{\bm \alpha_a}= \frac{\Phi^2}{2 A_{T^2}}$  where $A_{T^2}$ is the area of torus \cite{Banks:2008tpa}. Therefore, within BO  approximation, the reduced QM   possess  a unique perturbative vacuum, and 
  does not have instantons. It is not possible to obtain a knowledge concerning QFT from naive dimensional reduction with periodic compactification in BO-limit.   
  
  However, despite being correct, this is a little bit over-simplified.  
     For example, for $SU(2)$ gauge theory, the flux states are $ |0\rangle $,  and perturbatively degenerate $| \pm 1 \rangle,  | \pm 2 \rangle, \ldots $ pairs of states.  $| +1 \rangle$  mix up  with  $| -1 \rangle$ due to two instanton effects with action $2S_0$ where $S_0$ is the monopole action in QFT.   Therefore,   we can try  to  engineer a   vacuum structure by turning on background fluxes, such that the instantons of QFT survives in the ground state description of QM. 
     

 \vspace{0.3cm}
 \noindent
{\bf From Polyakov  to QM  (with  't Hooft  flux):} The  Polyakov  model  has a $\Z_N^{[1]}$  1-form symmetry, but no mixed anomalies.   We can turn on a discrete flux to examine the dynamics of the theory \cite{ tHooft:1979rtg, tHooft:1981nnx,  tHooft:1981sps,  vanBaal:1982ag, vanBaal:2000zc,   GarciaPerez:1992fj, GarciaPerez:1989gt, GonzalezArroyo:1995zy, GonzalezArroyo:1995ex, GarciaPerez:1993jw,Gonzalez-Arroyo:2019wpu, Kapustin:2014gua,  Gaiotto:2017yup, Unsal:2020yeh}. 
 Since the  model  is abelianized 
at long distances, we can replace our way of thinking in terms of discrete flux with a magnetic flux in co-weight lattice thanks to the relation $\Z_N \cong \Gamma_w^{\vee} / \Gamma_r^{\vee}  $.   
  Turning on the background flux,   $ {\bm  \mu}_1 \in \Gamma_w^{\vee}$, we end up with $N$ degenerate states, 
 \begin{align}
 | {\bm \nu}_1 \rangle \underbrace {\longrightarrow}_{    -{\bm  \alpha}_1}   | {\bm \nu}_2 \rangle   \underbrace {\longrightarrow}_{    -{\bm  \alpha}_2}  
 \cdots 
  \cdots  
    \underbrace {\longrightarrow}_{    -{\bm  \alpha}_{N-1}}    | {\bm \nu}_N \rangle  
           \label{tunnelilngsQM}
\end{align} 
connected to each other via monopole-events $\alpha_a \in \Delta^0$. Below, we argue that these instanton events have the same action as in $\R^3$. Assume $T^2$ size  $L$  obey 
\begin{align}
 r_{ \rm m} \ll L \ll d_{ \rm mm}
 \label{hier}
 \end{align}
  where $r_{ \rm m}$ is the monopole core size,  
and $d_{ \rm mm}$ is the characteristic distance between monopoles on $\R^3$.  This guarantees that   at a fixed Euclidean time slice $\tau$ and per $T^2$ size,   there will  typically be  at most one-monopole.  We also choose  $L \gg r_{ \rm m} $ so that the theory is locally   $3d$, and the action, which receives its contribution from the core region of monopole, remains  unchanged relative  to $\R^3$.   At distances $\tau  \gtrsim L$, the theory is correctly described by a simple quantum mechanical system with instantons, whose action are same as in the original 3d theory. 
 
 Here, we focus on $N=2$.   The insertion of 
$\Phi_{\rm bg} =\frac{1}{2}$ modifies the energetics of the set-up.  It  turns the flux states into $| n + \half \rangle$ with perturbative energies  $E_n =  { (n + \half )^2}/{2 A_{T^2}}$ and the fractional flux states $ |\pm  \half \rangle$ become degenerate. The tunnelling between them is an instanton effect with $\Delta \Phi = 1$ and action $S_0$, same as the instanton in full QFT. Note that the energy of the states  $| \pm \frac{3}{2} \rangle $ are higher, and in the BO approximation, we are justified to drop them, and SU(2) Polyakov with 't Hooft flux reduce to simple double-well potential. 


This story sounds  almost identical to particle on a circle in the presence of magnetic flux,  and a potential $-\cos(2q)$ leading to two harmonic minima.  At   $\theta = e \Phi_{\rm bg} = \pi$, all states are two-fold degenerate even non-perturbatively, 
because of mixed anomaly between $\Z_2$ translation symmetry  and time reversal $\rm T$  \cite{Gaiotto:2017yup,  Kikuchi:2017pcp}. 
In fact, 
$SU(2)$ Polyakov model  and $SU(2)$ deformed YM theories, reduce to double-well potential with configuration space $\R$ and  $S^1$ \cite{Unsal:2020yeh, Unsal:2021cch}, respectively.  The former does not have a mixed anomaly and the latter does.  The latter does and the two-fold degeneracy is   the remnant of  $\rm {CP}$  broken vacua of Yang-Mills theory in the QM limit.


In our reduced Polyakov model  with flux, the  two-fold degeneracy is lifted  non-perturnatively, and
  the ground and first excited states  are separated by a single monopole-instanton effect: 
	\begin{align}
    	|\Psi_0 \rangle  &= 
    		\textstyle {  \frac{1}{\sqrt 2} }\left( | \textstyle  \frac{1}{2}  \rangle  +  |  \textstyle { - \frac{1}{2} } \rangle \right),  \;  
		 	|\Psi_1 \rangle  =      		\textstyle {  \frac{1}{\sqrt 2} }\left( | \textstyle  \frac{1}{2}  \rangle  -  |  \textstyle { - \frac{1}{2} } \rangle \right),  \cr
\Delta E  &= 2 K e^{-S_0}	,
    	\end{align}
	where $S_0$ is monopole  action.  
	  Therefore, we claim that the resurgence properties in the QM limit of the Polyakov model  with flux is dictated by the  same instanton  action as  in $\R^3$.

In quantum mechanics,  for the double-well potential, the following resurgent cancellations are already proven 
by multiple methods   \cite{DDP, pham, Dunne:2014bca, Sueishi:2020rug}:
\begin{align}
&  \Im [{ \cal S}_{\pm}  P_0  +  [ {\cal M}_{\alpha_i} \overline {\cal M}_{\a_{i}}]_{\pm}  + \ldots ]=0  ,\cr
&   \Im \left[  [{\cal M }_{\a_{i}}]  \;  { \cal S}_{\pm}  P_{1}     +  [{\cal M }_{\a_{i}} \overline {\cal M}_{\a_{i} }{\cal M }_{\a_{i}}  ]_{\pm} + \ldots  \right] =0 .
  \label{res1}
\end{align}
Here,  
\begin{align}
P_0(g^2) =    \sum_{k=0}^{\infty}   {b}^{(0)}_k g^{2k}, \;\;\;   P_{1}  \left( g^2\right)   =  \sum_{k=0}^{\infty}   {b}^{(1)}_k g^{2k} 
\end{align}
are divergent asymptotic expansions around perturbative vacuum and 1-instanton saddle, and  
${ \cal S}_{\pm}$ indicate the lateral Borel resummations.  
 These series are non-Borel summable, i.e, the sum has an imaginary two-fold ambiguous part indicated by subscript $\pm$ in   ${ \cal S}_{\pm}$.   As described above, the 2-events and 3-events  in \eqref{res1} also have two-fold ambiguities. For the combination of the 
perturbation theory and the semi-classical analysis to be meaningful and ambiguity free,  these two types of ambiguities must cancel, and they do. 
In fact, exact WKB analysis and exact quantization conditions prove  that these cancellations and their generalizations hold true in all  non-perturbative sectors of QM \cite{pham, Sueishi:2020rug}. Note that  we work in the  $\lambda \rightarrow 0$ (BPS) limit, and investigate resurgent structure in $g^2$ only. This make sense provided $\lambda \ll g^2$.   At finite  $\lambda$, one needs a double-series and relatedly,  the action acquire a  $\lambda$ dependence,  $S_0= \frac{4 \pi}{g^2} f(\lambda)$   \cite{Kirkman:1981ck}.

  \vspace{0.3cm}
 \noindent
{\bf Back to  $\R^3$:}    Resurgent cancellations  \eqref{res1} are  not easy to prove in full QFT on $\R^3$.  But they are proven in the small   $T^2 \times \R$ 
with discretre flux,   a  construction in  which instantons of infinite volume theory survive. 
 In QFT, what we know rigorously is the  existence of ambiguity in the correlated events.  The imaginary ambiguous parts at the second and third order in  QFT on $\R^3$    are given by 
\begin{align}\label{Part1} 
&  \Im [ {\cal M}_{\alpha_i} \overline {\cal M}_{\a_{i}}]_{\pm}       \sim   \pm i   \left(\frac{{\rm s}_0}{g^2}\right)^{7}  e^{- 2{\rm s}_0/g^2} \cr 
&  \Im  [{\cal M }_{\a_{i}} \overline {\cal M}_{\a_{i} }{\cal M }_{\a_{i}}  ]_{\pm}   \sim  \pm i \left(\frac{{\rm s}_0}{g^2} \right)^{12} \ln \left(\frac{{\rm s}_0  }{g^2}  \right) \, e^{-\frac{3{\rm s}_0 }{g^2}}.
\end{align}
Here,     the power of   $ \left(\frac{{\rm s}_0}{g^2}\right)$ is 
$ 2 \nu + 3( \nu -1) $ for $\nu=2, 3, \ldots$   is the number of instantons that enter to correlated events. 
	 Recall that each monopole has 4 bosonic zero mode, and each zero mode induce $  ({{\rm s}_0}/{g^2} )^{1/2} $ in the prefactor, explaining $2 \nu$.  Each  quasi-zero mode direction gives a factor of three, hence $3( \nu -1) $. 
For a general $\nu$-instanton configuration, the power of  $\ln  \frac{{\rm s}_0  }{g^2}   $ is given by $\nu -2$.  

In order Polyakov's result on mass gap to be meaningful, (real, unambiguous), the counterpart of 
 \eqref{res1}  {\it must} hold in full QFT.  With this assumption,  and using dispersion relations such as 
$ b_k^{(0)} = \frac{1}{\pi}\int_0^\infty d (g^2)\, \frac{\Im  {\cal E}_{\rm np}   (g^2)}{(g^2)^{k+1}}  $, 
 we can determine the large-order growth of perturbation theory around perturbative vacuum and monopole saddle as: 
\begin{align}
 \label{pred-1}
b_k^{(0)} &\sim   \frac{\G(k+7)}{(2{\rm s}_0)^k}, \;\;\;   \quad  {\rm s}_0= 4\pi   \cr
b_k^{(1)} &\sim    \frac{\Gamma(k+10)\,  \ln (k+10) }{(2{\rm s}_0)^k} .
\end{align}	

%

Few remarks are in order.  Relative to standard quantum mechanical result for $b_k^{(0)}$ and  $b_k^{(1)}$, where  the factorial growth appears  generally as $\G(k+1)$ for ground state,  we obtain an enhancement. This is due to  the difference of the number of zero and quasi-zero modes in the two set-ups.  In $b_k^{(1)}$,  there is an extra  $\ln (k+10)$ enhancement as well. The log enhancement  also  appears in the context of quantum mechanics  around instanton sectors \cite{Alvarez-Howls,   Zinn-Justin:1983yiy, Dunne:2013ada}. It is there because  the 3-instanton 
amplitude  has a  $(\ln ( - {\rm s}_0/g^2))^2  $  in it, coming from integrating out of two quasi-zero modes, which leads to  log-dependent  imaginary part in \eqref{Part1}.   
It is quite  curious to note that $b_k^{(1)} \sim \frac{d}{d \nu} b_{k-2}^{(0)} |_{\nu=2} $ asymptotically,  reminiscent of  exact P/NP relation in quantum mechanics \cite{Dunne:2013ada, Alvarez}. The relation in QM is exact, it tells us that  perturbation theory around instanton is dictated by perturbation theory around perturbative vacuum via a simple relation.   It would  be  remarkable  if such a relation also holds in QFT. 
The  appearance of this  enhancement is  a relatively new effect in QFT, an example of which also appeared in \cite{Borinsky:2021hnd}.  But in retrospect, it is  inevitable, and generic. Application of stochastic perturbation theory on lattice can be useful to check these predictions \cite{DiRenzo:2004hhl,  DiRenzo:2008en, Gonzalez-Arroyo:2019zfm}, especially by modifying the scalar potential in  \cite{DiRenzo:2008en} to generate adjoint Higgsing and monopole confinement. 

%

  \vspace{0.3cm}
 \noindent
{\bf Conclusions:}
In order Polyakov's famous analysis for mass gap to be justified, one needs  \eqref{res1} to be true on $\R^3$.   We showed that the relation  \eqref{res1} is true in a special quantum mechanical reduction of Polyakov model with discrete  't Hooft  flux.  The reduction has the merit that it remembers the instanton of the theory on $\R^3$ on small $T^2 \times \R$ in the description of ground state.    
We emphasize that the use of 't Hooft flux   or  twisted boundary conditions   is not a nuisance  regardless of absence/presence  of a mixed anomaly. To the contrary, it is necessary  to make the instantons (or fractional instantons in deformed YM  on $\R^3 \times S^1$)   transparent in  a  quantum mechanical reduction \cite{ GarciaPerez:1992fj,  Gonzalez-Arroyo:2019wpu, Unsal:2020yeh, Cox:2021vsa}. 
 Assuming that     \eqref{res1}   continues to be valid  in the decompactification limit provides estimates of large-order structure of perturbation theory around perturbative vacuum and instanton \eqref{pred-1}.  We hope that these relations can be tested using stochastic perturbation theory, putting Polyakov's analysis on a firmer ground.

\vspace{0.3cm}
\noindent
{\bf Acknowledgments.}  M. \"U  thanks E. Witten for emphasizing the role of  critical points at infinity in an exchange in 2014, and  suggesting   that the bion effects can be viewed as emanating from them. 
We also thank Yuya Tanizaki, Sergei Gukov and Mendel Nguyen for discussions. 
M.\"U. acknowledges support from U.S. Department of Energy, Office of Science, Office of Nuclear Physics under Award Number DE-FG02-03ER41260.
C.P is supported by TUBITAK 2214-A Research Fellowship Programme for PhD Students.

\bibliography{QFT-Mithat}

\bibliographystyle{apsrev4-1}
\bibliographystyle{JHEP}

\end{document}